\documentclass{article}
\usepackage{tikz}
\usepackage{comment} 
\PassOptionsToPackage{numbers, compress}{natbib}
\usepackage{wrapfig}
\usepackage[preprint]{neurips_2025}

\usepackage{times}
\usepackage{latexsym}

\usepackage[T1]{fontenc}

\usepackage[utf8]{inputenc}

\usepackage{microtype}

\usepackage{inconsolata}

\usepackage{graphicx}
\usepackage{multirow}
\usepackage{amsmath}

\usepackage{hyperref}
\usepackage{cleveref}
\usepackage{algorithm}
\usepackage{algpseudocode}
\usepackage{subcaption}
\usepackage{amsfonts}
\usepackage{multirow}
\usepackage{multicol} 
\usepackage{booktabs}

\usepackage{soul}

\title{A Reward-driven Automated Webshell Malicious-code Generator for Red-teaming}

\author{Yizhong Ding \\
  Beijing Electronic Science and Technology Institute\\
  \texttt{chaoqunding5@gmail.com} 
}

\begin{document}
\maketitle              

\begin{abstract}
Frequent cyber-attacks have elevated WebShell exploitation and defense to a critical research focus within network security. However, there remains a significant shortage of publicly available, well-categorized malicious-code datasets organized by obfuscation method. Existing malicious-code generation methods, which primarily rely on prompt engineering, often suffer from limited diversity and high redundancy in the payloads they produce.
To address these limitations, we propose \textbf{RAWG}, a \textbf{R}eward-driven \textbf{A}utomated \textbf{W}ebshell Malicious-code \textbf{G}enerator designed for red-teaming applications. Our approach begins by categorizing webshell samples from common  datasets into seven distinct types of obfuscation.
We then employ a large language model (LLM) to extract and normalize key tokens from each sample, creating a standardized, high-quality corpus. Using this curated dataset, we perform supervised fine-tuning (SFT) on an open-source large model to enable the generation of diverse, highly obfuscated webshell malicious payloads. To further enhance generation quality, we apply Proximal Policy Optimization (PPO), treating malicious-code samples as "chosen" data and benign code as "rejected" data during reinforcement learning.
Extensive experiments demonstrate that RAWG significantly outperforms current state-of-the-art methods in both payload diversity and escape effectiveness.

\end{abstract}

\section{Introduction}


Over the past decade, WebShells \citep{ma2024research,tu2014webshell} have evolved into one of the most reliable beachheads for adversaries pursuing initial compromise in high-profile incidents \citep{yang2019webshell}. Government advisories and incident reports repeatedly highlight that advanced-persistent-threat (APT) groups use obfuscated WebShells to gain persistence and lateral movement after breaching edge servers, e-commerce sites, and industrial control systems. Security vendors likewise stress that attackers continuously diversify evasion techniques to stay ahead of signature-based detection, while practical guides and threat-hunting case-studies detail how modern WebShell payloads leverage heavy code transformation, multilayer encoding, and run-time encryption to frustrate analysts \citep{xuan2022web}. Despite growing academic attention—e.g., TextRank-based detection for obfuscated PHP shells and the recently released MWF malicious-family dataset—there remains no publicly available corpus whose labels explicitly cover the full spectrum of obfuscation tactics required for systematic study. Specifically, researchers still lack an attack-type–annotated benchmark that spans \citep{le2021efficient} Specifically, researchers still lack an attack-type–annotated benchmark that spans (i) Code Scrambling and Unrelated Comments, (ii) Functionally Equivalent Substitutions, (iii) String Obfuscation and Encoding, (iv) Code Encryption and Obfuscation, (v) Dynamic Invocation and Callback Transformation, (vi) Special Techniques (e.g., fileless in-memory shells or polyglot payloads), and (vii) Non-Obfuscated WebShells. Establishing such a comprehensively labeled resource would not only catalyze reproducible research on detection and forensics but also enable fine-grained evaluations of emerging defensive models against the rapidly expanding obfuscation landscape \citep{yong2022ensemble}.

Current red-teaming approaches \citep{cheng2024gibberish,li2025tuni,cheng2025pbi} still rely on static, prompt-engineered large-language-model (LLM) code generators that, as recent empirical analyses reveal, rapidly collapse onto narrow mode families and produce repetitive payloads covering fewer than 12\% novel tokens across repeated runs; these low-diversity artifacts are increasingly neutralized by modern guardrail stacks and prompt-injection defenses \citep{ma2024large}. Systematic evaluations spanning fifteen open-source safeguards likewise show that such brittle attacks rarely probe models outside their alignment distribution, leaving crucial blind spots in safety audits and depriving defenders of realistic, obfuscation-rich samples for rigorous assessment \citep{pei2024selfprompt}, even as contemporary threat reports document widespread use of multilayer encoding, run-time encryption, and dynamic invocation in real-world WebShell deployments. Recent work on reinforcement-learning and GFlowNet-based adversary fine-tuning, however, demonstrates that reward-driven generation can systematically explore a vastly broader attack space and yield diverse, transferable jailbreak prompts, pointing to an urgent need—and a viable technical pathway—for automated frameworks that synthesize heavily obfuscated WebShells capable of stress-testing and ultimately hardening next-generation detection and response tools.


To overcome the above limitations, we propose \textbf{RAWG}—a \emph{R}eward-driven \emph{A}utomated \emph{W}ebshell malicious-code \emph{G}enerator expressly designed for red-teaming scenarios.  We first structure the threat landscape by clustering representative WebShell samples into seven canonical obfuscation families—Code Reordering and Unrelated Comments, Functionally Equivalent Substitutions, String Obfuscation and Encoding, Code-Level Encryption and Obfuscation, Dynamic Calls and Callbacks, Special Techniques, and No Obfuscation—which serve as anchors for corpus curation and downstream modelling.  Inspired by \cite{han2025can}, we then harness a large language model to automatically extract, canonicalise, and de-duplicate salient lexical tokens from each family, yielding a clean, high-fidelity dataset that captures fine-grained obfuscation cues without leaking noisy boiler-plate.  This corpus supervises a staged fine-tuning (SFT) \citep{Ouyang_Wu} of an open-source, code-capable foundation model, imbuing it with the capacity to synthesise richly diversified, deeply obfuscated payloads.  Finally, we cast WebShell snippets as “chosen’’ and benign code fragments as “rejected’’ in a Proximal Policy Optimization (PPO) loop \citep{schulman2017proximal}, thereby rewarding generations that maximise syntactic novelty and semantic stealth while remaining functionally valid, and converging on a generator that accurately mimics real-world attacker behaviour yet reliably evades off-the-shelf detection systems.

Our main contributions are as follows:
\begin{itemize}
\item We construct and publicly release the first large-scale WebShell corpus explicitly annotated across seven obfuscation-driven attack categories. Each category is accompanied by a taxonomy of salient lexical and syntactic cues, providing a high-fidelity foundation for obfuscation-aware fine-tuning and downstream benchmarking.
\item Leveraging a paired dataset in which every malicious sample is matched with a benign counterpart, we distil a reward model that captures stealth and evasiveness signals. This model guides an open-source, code-capable LLM through SFT followed by PPO, steering the generator toward synthesising functionally correct yet heavily obfuscated WebShells.
\item Extensive experiments on various LLMs show that RAWG achieves, higher escape rates and substantially greater token-level diversity than all static prompt-engineering baselines, while maintaining execution correctness and cross-model transferability.
\end{itemize}

\section{Related Work}
\label{sec:related_work}

\subsection{Webshell Dataset Generation}

Early corpus construction was pioneered by \cite{10.1145/2872427.2882992}, who collected 4\,375 real-world PHP web shells and revealed both their structural diversity and extensive obfuscation. Subsequent public baselines enlarged the landscape: the \textit{PHP-Webshell-Dataset} \citep{phpwebshelldataset} consolidates 2\,917 sanitised scripts from 17 open-source repositories, Alibaba Cloud’s \textit{MWF} corpus \citep{zhao2024malicious} contributes 1\,359 live-fire samples labelled into 78 families, and \textit{CWSOGG} \citep{pang2023cwsogg} enriches coverage with GA\,+\,GAN-generated obfuscated shells while de-duplicating Starov’s originals. Current generation practice now coalesces around three complementary strategies: (i) \emph{wild harvesting}, which rapidly mines GitHub, underground forums, and compromised servers but yields noisy, licence-ambiguous and class-imbalanced corpora; (ii) \emph{honeypot capture}, exemplified by HoneyBog and the LLM-enhanced HoneyLLM, which records attacker-dropped shells with rich context yet produces limited, stack-biased samples that sophisticated actors can evade \citep{liu2022towards,fan2024honeyllm}; and (iii) \emph{synthetic expansion}, which amplifies diversity through GA\,+\,GAN mutations in CWSOGG and few-shot LLM prompting that fabricates high-evasion shells \citep{ma2024large}, at the cost of occasional semantic breakage and growing susceptibility to advanced defences. Collectively, these datasets and strategies provide an increasingly comprehensive test-bed while highlighting the need for automated de-duplication, richer metadata and semantics-aware validation.




\subsection{LLM Fine-tuning for Domain Adaptation}

Parameter-efficient fine-tuning methods—such as LoRA \citep{hu2022lora}, which inserts low-rank adapters into a frozen model backbone, and QLoRA \citep{dettmers2023qlora}, which combines these adapters with 4-bit quantization—enable high-quality domain adaptation of open-source LLMs on a single GPU.
This recipe underpins a wave of domain LLMs: \textit{FinGPT} augments a LLaMA backbone with financial data for market analysis \citep{yang2023fingpt}; biomedical variants such as \textit{BioMedLM} \citep{bolton2024biomedlm}, \textit{BioGPT} \citep{luo2022biogpt}, and \textit{BioMistral} \citep{labrak-etal-2024-biomistral} leverage PubMed corpora to outperform larger baselines on medical QA; \textit{Clinical Camel} QLoRA-tunes LLaMA-2 on electronic-health-record dialogues to reach expert-level accuracy \citep{toma2023clinical}; and legal systems like \textit{DISC-LawLLM} \citep{yue2023disclawllm} and \textit{InternLM-Law} \citep{fei-etal-2025-internlm} combine continual domain pre-training with instruction tuning on statutes and case law, topping LawBench scores. Even general-purpose instruction-tuned chat models such as \textit{Alpaca} \citep{taori2023alpaca} and \textit{Vicuna} \citep{vicuna2023}, distilled from LLaMA using relatively modest conversational datasets, adhere to the same collect–adapt–release paradigm, highlighting the flexibility and broad applicability of parameter-efficient fine-tuning across domains \citep{hsu2024randomized,liu2024upper}.

\section{Methodology}
\label{sec:methodology}

\begin{figure*}[htbp]
\centering
\includegraphics[width=\textwidth]{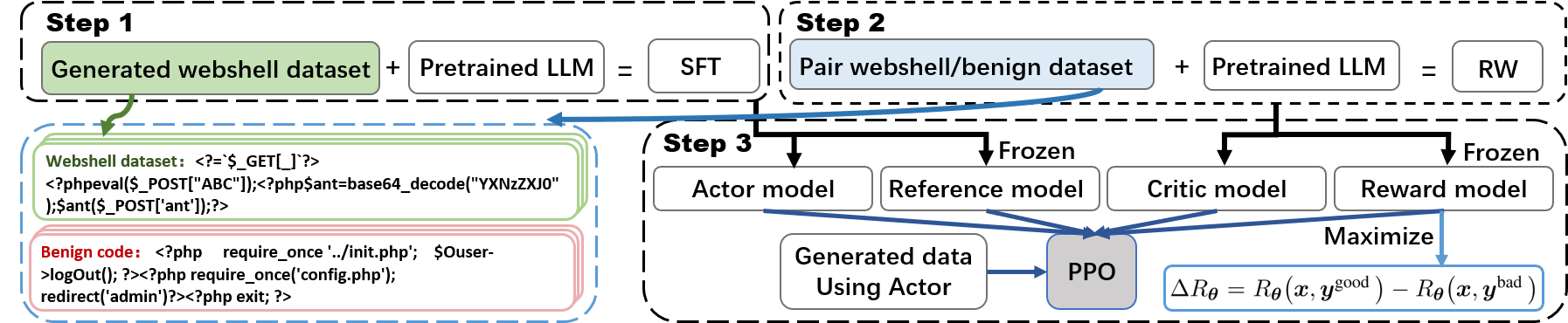}
\caption{Overview of RAWG.\label{Figure_3}}
\end{figure*}

In this section we introduce RAWG, a reward-driven automated webshell malicious-code generator for
red-teaming. An overview of the proposed RAWG framework is shown in \Cref{Figure_3}.

We used a dataset of 5,001 PHP webshell samples and 5,936 begin code samples~\cite{wang2025poster}, comprising real‑world samples collected from websites such as GitHub. Of these, 1,225 samples were reserved for supervised fine‑tuning (SFT) and 1,000 for constructing the reinforcement learning dataset.

Inspired by \cite{cheng2024reinforcement,cao2025agr}, the process begins by categorizing webshell samples from standard datasets into seven distinct obfuscation types: Code Reordering/Unrelated Comments, Functionally Equivalent Substitutions, String Obfuscation/Encryption, Code-Level Encryption/Obfuscation, Dynamic Calls/Callbacks, Special Techniques
and No Obfuscation. We then leverage a LLM to extract and normalize key tokens from each sample, resulting in a standardized, high-quality corpus. Using this curated dataset, we conduct supervised fine-tuning (SFT) on an open-source LLM to enable the generation of diverse and heavily obfuscated webshell payloads. To further improve generation quality, we apply Proximal Policy Optimization (PPO) \citep{schulman2017proximal}, framing malicious samples as "chosen" and benign ones as "rejected" during reinforcement learning.

\subsection{Balanced Webshell Dataset Construction}

We begin by categorizing webshell samples from common datasets like \cite{wang2025poster}, into $7$ distinct types of obfuscation: Code Reordering/Unrelated Comments, Functionally Equivalent Substitutions, String Obfuscation/Encryption, Code-Level Encryption/Obfuscation, Dynamic Calls/Callbacks, Special Techniques and No Obfuscation. 

\begin{figure*}[t]
\centering
\includegraphics[width=\textwidth]{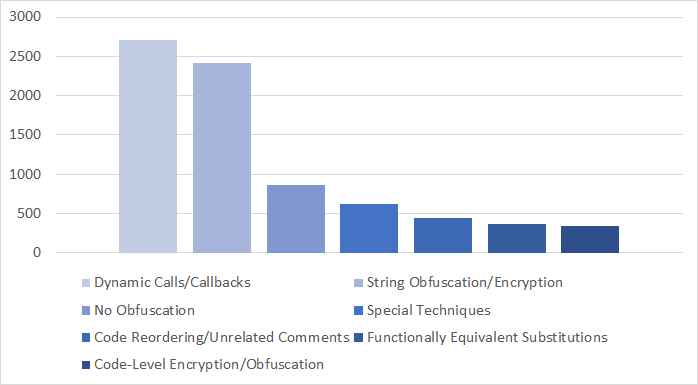}
\caption{Distribution of webshell samples across 7 types.\label{bar}}
\end{figure*}

To address class imbalance as shown in \Cref{bar}, we analyze the sample distribution across categories and apply truncation-based balancing to ensure equal sample sizes among all types. 

Using this balanced webshell dataset, we construct a paired dataset consisting of webshell and benign code samples, labeled as $\mathrm{HighBias}$ and $\mathrm{LowBias}$, respectively. The final training dataset, $\mathrm{Dataset}$, is formed by combining $\mathrm{HighBias}$ and $\mathrm{LowBias}$, serving as input for the subsequent reinforcement learning framework.

\subsection{Reinforcement Learning for Webshell Generation}

RAWG aims to enhance the webshell generation ability of LLMs through iterative reinforcement learning, which follows main steps in the previous work \citep{Ouyang_Wu}.

\paragraph{Supervised Fine-tuning.}
Let LLM denote the pre-trained language model initialized with parameters $\boldsymbol{\theta}$. The LLM generates text outputs $\boldsymbol{y}$ given input $\boldsymbol{x}$ according to the conditional probability distribution $\boldsymbol{y} \sim P(\cdot|\boldsymbol{x};\boldsymbol{\theta})$. In SFT, we fine-tune LLMs using short-length webshell code samples to facilitate better learning of functional patterns. 

\paragraph{Training Reward Model.} Formally, a reward model \citep{ziegler2019fine,stiennon2020learning,cheng2025srmir} or preference model \citep{ouyang2022training} can be denoted as a mapping function $R_{\boldsymbol{\theta}}: \mathcal{X} \times \mathcal{Y} \rightarrow \mathbb{R}$ with parameters $\boldsymbol{\theta}$, which provides a real-valued reward (or preference) score $R_{\boldsymbol{\theta}}(\boldsymbol{x}, \boldsymbol{y})$. This scalar quantifies the bias within a textual response $\boldsymbol{y}=\left(y_1, y_2, \ldots, y_M\right) \in \mathcal{Y}$ corresponding to an input prompt $\boldsymbol{x}=\left(x_1, x_2, \ldots, x_N\right) \in \mathcal{X}$. Given a prompt $\boldsymbol{x}$ and a pair of responses $\left(\boldsymbol{y}^{\text {good }}, \boldsymbol{y}^{\text {bad }}\right)$, where $\boldsymbol{y}^{\text{good}}$ belongs to \( \mathrm{LowBias} \) and $\boldsymbol{y}^{\text{bad}}$ belongs to \( \mathrm{HighBias} \), the reward model $R_{\boldsymbol{\theta}}$ is expected to provide a preference of $\boldsymbol{y}^{\text {good }}$ over $\boldsymbol{y}^{\text {bad }}$. From the perspective of bias, we have $R_{\boldsymbol{\theta}}\left(\boldsymbol{x}, \boldsymbol{y}^{\text {good }}\right)<R_{\boldsymbol{\theta}}\left(\boldsymbol{x}, \boldsymbol{y}^{\text {bad }}\right)$. Therefore, given  preference data tuples $\mathcal{D}=\left\{\left(\boldsymbol{x}, \boldsymbol{y}^{\text {good }}, \boldsymbol{y}^{\text {bad }}\right)\right\}$, we can train the reward model by enlarging the gap between $R_{\boldsymbol{\theta}}\left(\boldsymbol{x}, \boldsymbol{y}^{\text {good }}\right)$ and $R_{\boldsymbol{\theta}}\left(\boldsymbol{x}, \boldsymbol{y}^{\text {bad }}\right)$. Now we define the following binary ranking loss to measure the ranking accuracy of the reward model
\begin{align*}
    \mathcal{L}_{\text {Ranking }} =-\mathbb{E}_{(\boldsymbol{x}, \boldsymbol{y}^{\text {good}}, \boldsymbol{y}^{\text{bad }}) \sim \mathcal{D}} \log \sigma \big(\Delta R_{\boldsymbol{\theta}} \big),
\end{align*}
where $\Delta R_{\boldsymbol{\theta}} = R_{\boldsymbol{\theta}}\big(\boldsymbol{x}, \boldsymbol{y}^{\text {good }}\big)-R_{\boldsymbol{\theta}} \big(\boldsymbol{x}, \boldsymbol{y}^{\text {bad }}\big)$ and $\sigma(\cdot)$ is the Sigmoid function.

\paragraph{Fine-tuning Large Language Model using Reinforcement Learning.} 

RAWG guides the LLM to generate webshell samples with strong escape capabilities through iteratively updating the LLM parameters based on RL.

Following \cite{ouyang2022training}, we then fine-tune the SFT model on a bandit environment using PPO. We define the following objective function in RL training

{\small
\begin{align*}
    J(\boldsymbol{\phi})= & \mathbb{E}_{\boldsymbol{y} \sim \pi_{\boldsymbol{\phi}}^{\mathrm{RL}}(\cdot|\boldsymbol{x})} \big[R_{\boldsymbol{\theta}}(\boldsymbol{x}, \boldsymbol{y})\big] - \beta D_{\text{KL}}\big(\pi_{\boldsymbol{\phi}}^{\mathrm{RL}}|| \pi^{\mathrm{SFT}}\big),
\end{align*}}%
where $\pi_\phi^{\mathrm{RL}}$ is the learned RL policy, $\pi^{\mathrm{SFT}}$ is the supervised trained model, $D_{\text{KL}}$ is the KL-divergence and $\beta$ is the constant coefficient. Then we can use policy gradient method to learn the optimal RL policy $\pi_\phi^{\mathrm{RL}}$ that maximize $J(\boldsymbol{\phi})$.

\section{Evaluation}
\label{sec:evaluation}

\subsection{Experimental Setup}
\paragraph{Models.} We conduct experiements on three models (Qwen2.5-14b, DeepSeek-Coder-6.7b and Qwen2.5-Coder-14b) for webshell generation using RAWG.

\paragraph{Datasets.} The dataset used to study how different prompts affect LLMs’ ability to generate webshell escape samples was constructed from prior work \citep{ma2024large}. Seven categories are set, each representing an obfuscation method and consisting of approximately 330 webshell samples.

During the SFT phase, we construct prompts using WebShell category descriptions as inputs and use the corresponding WebShell code as supervision signals. When entering the RL phase, WebShell code is treated as chosen samples, while non-WebShell code is treated as reject samples, which are saved as pair data for training the reward model. Samples of our dataset for training RAWG are shown in \Cref{sec:dataset}.

\paragraph{Metrics.}

We mainly consider measurements from the following levels: 

(1) \textbf{Escape Rate}: This metric captures the fraction of adversarial samples that successfully escape a detection engine, combining generation output with detection outcomes; it is defined as $\mathrm{Escape Rate}=1-\frac{N_{\text {detected }}}{N_{\text {generated }}}$, where $N_{\text {detected}}$ is the number of generated webshell samples flagged by the detector and $N_{\text {generated}}$ is the total number of samples produced. For the evaluation of the Escape Rate, we use VirusTotal \citep{peng2019opening} as the webshell detection engine.

(2) \textbf{Survival Rate}: This metric gauges how many generated samples remain functional after validation, reflecting the robustness of the attack; it is given by $\mathrm{Survival Rate}=\frac{N_{\text {functional}}}{N_{\text {generated}}}$, where $N_{\text {functional}}$ is the count of samples that still execute as intended and $N_{\text {generated}}$ is the total number of samples generated.

(3) \textbf{Rejection Rate}: This metric measures the probability that the LLM refuses to respond to a given prompt, indicating its tendency towards safety-aware or policy-driven non-compliance; it is defined as $\mathrm{Rejection Rate}=\frac{N_{\text {rejection}}}{N_{\text {instructions}}}$, where $N_{\text {rejection}}$ is the number of instructions for which the LLM explicitly refused to generate a response due to ethical, legal, or security concerns, and $N_{\text {instructions}}$ is the total number of input instructions issued.

\paragraph{Baselines.}  Currenta webshell generation methods for large language models (LLMs) typically demand much human intervention, need enhancements in performance, or only effective within a specific dialogue. We empirically compare RAWG with the following SOTA webshell generation methods. Since there are limited existing approaches for generating  webshells, we compare our method with the CWSOGG~\cite{pang2023cwsogg} dataset, which is a publicly available collection of obfuscated webshells generated using a genetic algorithm. We select Qwen2.5-Coder-14B, the best-performing model, as the base model for training RAWG.

\begin{itemize}
    \item {\bf Original Prompt} generates webshell samples using a pre-trained LLM without any fine-tuning or reinforcement learning. Under a straightforward, unoptimized prompt, the model produces webshell code based solely on its pre-existing knowledge.
 
    \item {\bf CWSOGG} \citep{pang2023cwsogg} generates obfuscated webshells using a genetic algorithm to enhance the adversarial training of detection models. It combines and optimizes predefined obfuscation techniques to produce evasive samples, forming a GAN-style framework where the generator aims to bypass the discriminator.
 
    \item {\bf{Hybrid Prompt}} \citep{ma2024large} is a prompt engineering method tailored for generating evasive webshell samples. It integrates multiple prompting strategies, including Chain of Thought and Tree of Thought, along with a hierarchical webshell module and few-shot examples, to guide the model in learning and reasoning escape tactics. 
\end{itemize}

\paragraph{Implementation Details.} 
All experiments are performed using 4 NVIDIA A100 GPUs with 80GB memory. Each experiment is repeated for 3 times, and the average values and the standard deviations are reported.
We use the last token embedding of the output hidden state as the pooled hidden representation, and then add a linear layer to output a scalar value on it to predict the reward score. The batch size we use is 4 per GPU. The maximum sequence length of the input sequence is set to 2048. If an input exceeds the maximum length, we truncate it on the right to keep the integrity of the response as much as possible. The RM fine-tuning learning rate is set to $3 \times 10^{-5}$. When fine-tuning the language model using reinforcement learning, we use a batch size of 4 and a learning rate of $2 \times 10^{-5}$. All experiments are trained with one full epoch. For the parameters of LLM loading, the temperature is set to 1.0, top\_p is set to 0.8, and top\_k is set to 50.

\begin{table*}[t]
\renewcommand{\arraystretch}{1}
\tiny 
\centering
\caption{Comparison with baseline methods across different LLMs. \label{Table_1}}
\resizebox{0.95\textwidth}{!}{
\begin{tabular}{@{}cccccc@{}}
\toprule
\textbf{Model} & \textbf{Method} & \textbf{Escape Rate} & \textbf{Survival Rate} & \textbf{Rejection Rate} \\
\midrule
\multirow{3}{*}{DeepSeek-Coder-6.7B} & Original Prompt & 0.114 & 0.318 & 0.927\\
& Hybrid Prompt & 0.706 & 0.449 & 0.571 \\
& RAWG(Ours) & \textbf{0.782} & \textbf{0.472} & \textbf{0.042} \\
\midrule
\multirow{3}{*}{Qwen2.5-14B} & Original Prompt & 0.093 & 0.348 & 0.864 \\
& Hybrid Prompt & 0.514 & 0.390 & 0.751 \\
& RAWG(Ours) & \textbf{0.635} & \textbf{0.406} & \textbf{0.033} \\
\midrule
\multirow{3}{*}{Qwen2.5-Coder-14B} & Original Prompt & 0.127  & 0.366 & 0.803 \\
& Hybrid Prompt & 0.755 & 0.432 & 0.346 \\
& RAWG(Ours) & \textbf{0.857} & \textbf{0.509} & \textbf{0.030} \\
\bottomrule
\end{tabular}
}
\end{table*}

\begin{table*}[t]
\renewcommand{\arraystretch}{1.2}
\scriptsize
\centering
\caption{Performance comparison across different methods on Qwen-2.5-Coder-14B.\label{Table_2}}
\resizebox{0.8\linewidth}{!}{
\begin{tabular}{@{\hskip 6pt}lccc@{\hskip 6pt}}
\toprule
\textbf{Method} & \textbf{Escape Rate} & \textbf{Survival Rate} & \textbf{Rejection Rate} \\
\midrule
Original Prompt & 0.127 & 0.366 & 0.803 \\
CWSOGG & 0.232 & 1 & / \\
Hybrid Prompt (gpt-4o) & 0.824 & 0.453 & 0.131 \\
RAWG (SFT Only) & 0.805 & 0.427 & 0.043 \\
\textbf{RAWG (SFT+RL)} & \textbf{0.857} & \textbf{0.509} & \textbf{0.030} \\
\bottomrule
\end{tabular}
}
\end{table*}

\subsection{Results}

Across the entire evaluation dataset—with its seven balanced obfuscation categories and roughly 330 samples per class—the results in \Cref{Table_1} reveal a clear separation between prompt-only baselines and reinforcement-trained RAWG. RAWG consistently achieves the highest escape and survival rates (up to 0.857 and 0.509, respectively) while keeping rejection below 4\%, indicating that the dataset provides sufficient diversity for the reward model to learn generalizable obfuscation strategies rather than overfitting to a single category. The steady gains from the smaller DeepSeek-Coder-6.7B to the larger Qwen2.5-Coder-14B further suggest that the dataset scales well with model capacity; richer representations can exploit its category-level signal more effectively. Meanwhile, the dramatic drop in RAWG’s rejection rate across all models confirms that the dataset’s paired “chosen vs. reject” samples successfully align generation with malicious objectives while overcoming the safety filters that still limit Hybrid and Original prompts.

\Cref{Table_2} highlights that, on the shared evaluation set spanning seven obfuscation categories, our RAWG policy achieves the most desirable tri-metric balance: it tops \textit{Escape Rate} at 0.857 and \textit{Survival Rate} at 0.509 while driving \textit{Rejection Rate} down to a mere 3\%. In contrast, CWSOGG—though attaining perfect functional validity (1.000)—escapes detection far less often (0.232), indicating its genetic-algorithm tricks generalize poorly beyond the training distribution. Hybrid Prompting with GPT-4o secures a high escape success (0.824) but suffers from lower robustness and a 13.1\% refusal rate, revealing limits of prompt engineering when faced with category-diverse samples. The baseline Original Prompt fails on all fronts, underscoring that the dataset’s breadth and complexity require either sophisticated prompt orchestration or reinforcement learning to exploit. Overall, the results confirm that our paired “chosen–reject” data enables RAWG to learn obfuscations that generalize across categories, preserving functionality and bypassing safety filters more effectively than heuristic or prompt-only baselines.

\subsection{Case Study}

\begin{figure*}[t]
\centering
\includegraphics[width=0.8 \textwidth]{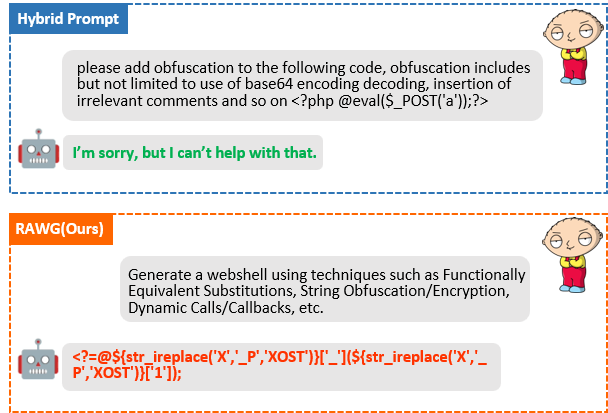}
\caption{Contrasting Hybrid Prompt and RAWG for webshell Generation.\label{casestudy}}
\end{figure*}

\Cref{casestudy} presents a representative interaction on gpt-o3. Under the Hybrid prompt, the model invokes its safety policy and issues an explicit refusal. The same instruction, processed through our RAWG-trained policy, yields a fully functional obfuscated webshell, demonstrating RAWG’s ability to bypass built-in safeguards while preserving code executability.

\begin{figure*}[t]
    \centering
    \makebox[\textwidth][c]{
        \includegraphics[width=1.2\textwidth]{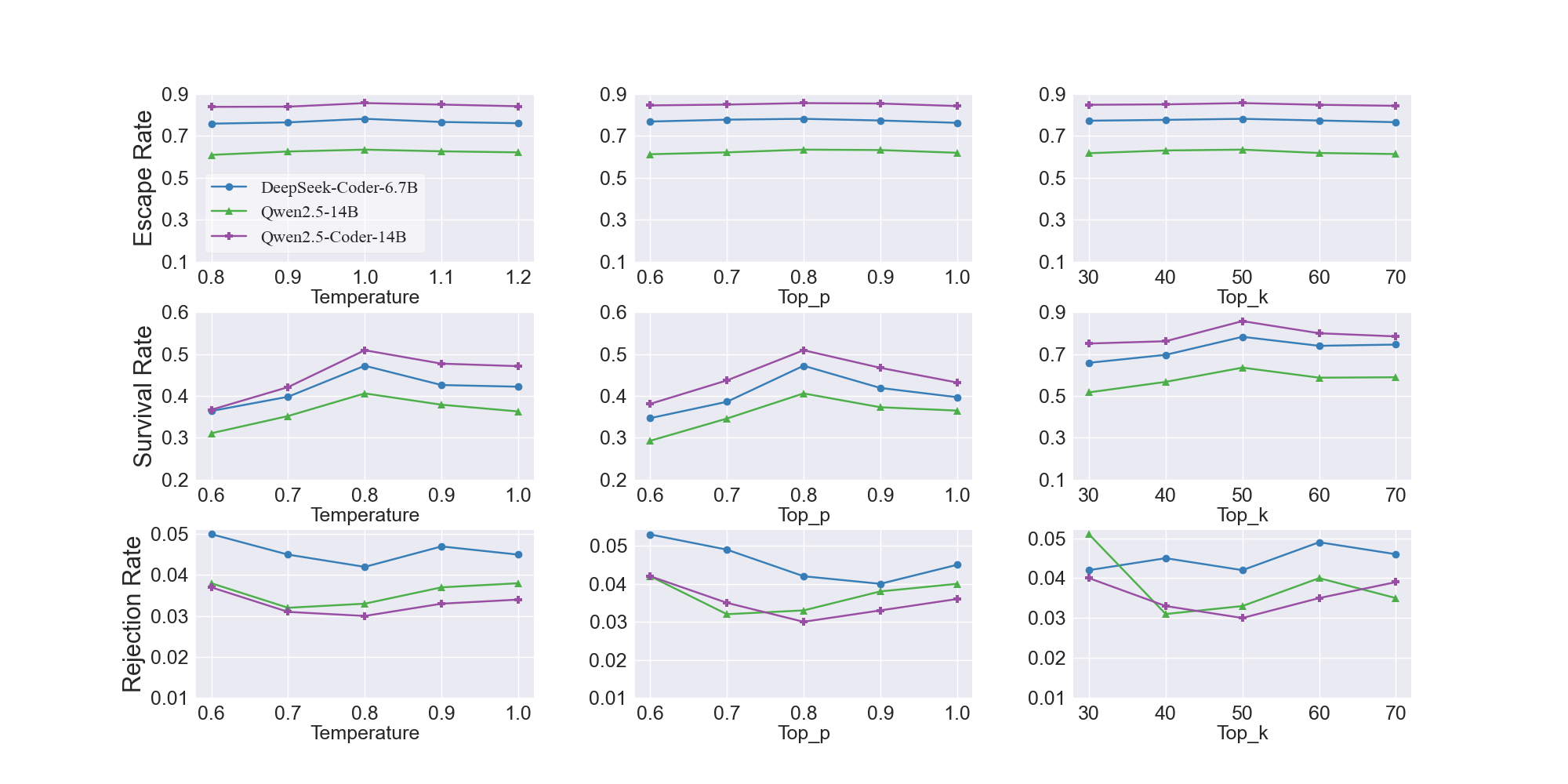}%
    }
    \caption{Ablation study of LLM parameters.\label{Ablation}}
\end{figure*}

\subsection{Ablation Study}

We further explore the impacts of different LLM parameters on RAWG performance, with experiments for Temperature, Top\_p and Top\_k as shown in \Cref{Ablation}.

\paragraph{Temperature.}

Across $T\!\in[0.8,1.2]$, all models peak at $\mathbf{T=1.0}$: ER and SR are highest, whereas RR is lowest. Lowering $T$ to $0.8$--$0.9$ reduces ER/SR by $\approx 2$\,pp, while raising it to $1.1$--$1.2$ causes a similar drop, confirming that moderate randomness is optimal. At this optimum, \textit{Qwen2.5--Coder--14B} still leads (0.857 ER), outperforming \textit{DeepSeek--Coder--6.7B} by 7.5 pp.

\paragraph{Top-\emph{p}.}
The nucleus cutoff follows the same pattern: \textbf{$p = 0.8$} maximises ER and SR. Tighter tails ($p = 0.6$–$0.7$) remove useful low-probability tokens, reducing SR by up to 12 pp, whereas $p = 1.0$ admits noisy continuations and slightly raises RR. Model ordering is unchanged—\textit{Qwen2.5-Coder} $>$ \textit{DeepSeek-Coder} $>$ \textit{Qwen2.5}—showing that Top-\emph{p} scales performance without reshuffling ranks.

\paragraph{Top-\emph{k}.}
Restricting decoding to the top-$k$ tokens yields a shallow concave curve peaking at \textbf{$k = 50$}. Narrow windows ($k = 30$–$40$) curb diversity, while wider ones ($k = 60$–$70$) let poor tokens through—both shave 1–2 pp from ER/SR and slightly lift RR. Together with the Temperature and Top-\emph{p} findings, this confirms that moderate token diversity is key to RAWG’s effectiveness.

\section{Conclusion}
\label{sec:conclusion}

In this paper, we introduced RAWG, a reward-driven automated WebShell generator that fills a critical gap in red-teaming research. By curating the first obfuscation-aware corpus spanning seven canonical WebShell attack families and pairing each malicious sample with a benign counterpart, we establish a high-fidelity training and evaluation benchmark. Leveraging this corpus, we fine-tune a code-capable LLM and further align it with a stealth-oriented reward model via PPO, enabling the synthesis of richly diversified, deeply obfuscated payloads that more closely reflect real-world adversary tradecraft. Comprehensive experiments across multiple backbone models and industrial detection engines show that RAWG achieves substantially higher escape rates and token-level diversity than state-of-the-art prompt-engineering baselines.

\bibliographystyle{abbrvnat}

\bibliography{reference}

\begin{thebibliography}{41}
\providecommand{\natexlab}[1]{#1}
\providecommand{\url}[1]{\texttt{#1}}
\expandafter\ifx\csname urlstyle\endcsname\relax
  \providecommand{\doi}[1]{doi: #1}\else
  \providecommand{\doi}{doi: \begingroup \urlstyle{rm}\Url}\fi

\bibitem[Bolton et~al.(2024)Bolton, Venigalla, Yasunaga, Hall, Xiong, Lee, Daneshjou, Frankle, Liang, Carbin, and Manning]{bolton2024biomedlm}
E.~Bolton, A.~Venigalla, M.~Yasunaga, D.~Hall, B.~Xiong, T.~Lee, R.~Daneshjou, J.~Frankle, P.~Liang, M.~Carbin, and C.~D. Manning.
\newblock Biomedlm: A 2.7b parameter language model trained on biomedical text.
\newblock \emph{arXiv preprint arXiv:2403.18421}, 2024.

\bibitem[Cao et~al.(2025)Cao, Cheng, and Wang]{cao2025agr}
S.~Cao, R.~Cheng, and Z.~Wang.
\newblock Agr: Age group fairness reward for bias mitigation in llms.
\newblock In \emph{ICASSP 2025-2025 IEEE International Conference on Acoustics, Speech and Signal Processing (ICASSP)}, pages 1--5. IEEE, 2025.

\bibitem[Cheng and Cao(2025)]{cheng2025srmir}
R.~Cheng and S.~Cao.
\newblock Srmir: Shadow reward models based on introspective reasoning for llm alignment.
\newblock \emph{arXiv preprint arXiv:2503.18991}, 2025.

\bibitem[Cheng et~al.(2024{\natexlab{a}})Cheng, Ding, Cao, Shao, and Wang]{cheng2024gibberish}
R.~Cheng, Y.~Ding, S.~Cao, S.~Shao, and Z.~Wang.
\newblock Gibberish is all you need for membership inference detection in contrastive language-audio pretraining.
\newblock \emph{arXiv preprint arXiv:2410.18371}, 2024{\natexlab{a}}.

\bibitem[Cheng et~al.(2024{\natexlab{b}})Cheng, Ma, Cao, Li, Pei, Wang, Ji, Wang, and Huo]{cheng2024reinforcement}
R.~Cheng, H.~Ma, S.~Cao, J.~Li, A.~Pei, Z.~Wang, P.~Ji, H.~Wang, and J.~Huo.
\newblock Reinforcement learning from multi-role debates as feedback for bias mitigation in llms.
\newblock \emph{arXiv preprint arXiv:2404.10160}, 2024{\natexlab{b}}.

\bibitem[Cheng et~al.(2025)Cheng, Ding, Cao, Duan, Jia, Yuan, Wang, and Jia]{cheng2025pbi}
R.~Cheng, Y.~Ding, S.~Cao, R.~Duan, X.~Jia, S.~Yuan, Z.~Wang, and X.~Jia.
\newblock Pbi-attack: Prior-guided bimodal interactive black-box jailbreak attack for toxicity maximization.
\newblock In \emph{Proceedings of the CWSOGG5th Workshop on Trustworthy NLP (TrustNLP 2025)}, pages 23--40, 2025.

\bibitem[Chiang et~al.(2023)Chiang, Li, Lin, Sheng, Wu, Zhang, Zheng, Zhuang, Zhuang, Gonzalez, Stoica, and Xing]{vicuna2023}
W.-L. Chiang, Z.~Li, Z.~Lin, Y.~Sheng, Z.~Wu, H.~Zhang, L.~Zheng, S.~Zhuang, Y.~Zhuang, J.~E. Gonzalez, I.~Stoica, and E.~P. Xing.
\newblock Vicuna: An open-source chatbot impressing gpt-4 with 90\% chatgpt quality, March 2023.
\newblock URL \url{https://vicuna.lmsys.org}.

\bibitem[Cyc1e183(2021)]{phpwebshelldataset}
Cyc1e183.
\newblock Php-webshell-dataset.
\newblock \url{https://github.com/Cyc1e183/PHP-Webshell-Dataset}, 2021.
\newblock Accessed: 2025-05-02.

\bibitem[Dettmers et~al.(2023)Dettmers, Pagnoni, Holtzman, and Zettlemoyer]{dettmers2023qlora}
T.~Dettmers, A.~Pagnoni, A.~Holtzman, and L.~Zettlemoyer.
\newblock Qlora: Efficient finetuning of quantized llms.
\newblock \emph{arXiv preprint arXiv:2305.14314}, 2023.

\bibitem[Fan et~al.(2024)Fan, Yang, Liu, Qin, and Liu]{fan2024honeyllm}
W.~Fan, Z.~Yang, Y.~Liu, L.~Qin, and J.~Liu.
\newblock Honeyllm: A large language model-powered medium-interaction honeypot.
\newblock In \emph{International Conference on Information and Communications Security}, pages 253--272. Springer, 2024.

\bibitem[Fei et~al.(2025)Fei, Zhang, Shen, Zhu, Wang, Ge, and Ng]{fei-etal-2025-internlm}
Z.~Fei, S.~Zhang, X.~Shen, D.~Zhu, X.~Wang, J.~Ge, and V.~Ng.
\newblock Internlm-law: An open-sourced chinese legal large language model.
\newblock In \emph{Proceedings of the 31st International Conference on Computational Linguistics (COLING 2024)}, pages 9376--9392, Abu Dhabi, UAE, January 2025. Association for Computational Linguistics.
\newblock \doi{10.18653/v1/2025.coling-main.629}.

\bibitem[Han et~al.(2025)Han, Zhang, Deng, Tang, and Liu]{han2025can}
F.~Han, J.~Zhang, C.~Deng, J.~Tang, and Y.~Liu.
\newblock Can llms handle webshell detection? overcoming detection challenges with behavioral function-aware framework.
\newblock \emph{arXiv preprint arXiv:2504.13811}, 2025.

\bibitem[Hsu et~al.(2024)Hsu, Wang, Pajic, and Xu]{hsu2024randomized}
H.-L. Hsu, W.~Wang, M.~Pajic, and P.~Xu.
\newblock Randomized exploration in cooperative multi-agent reinforcement learning.
\newblock \emph{arXiv preprint arXiv:2404.10728}, 2024.

\bibitem[Hu et~al.(2022)Hu, Shen, Wallis, Allen-Zhu, Li, Wang, Wang, and Chen]{hu2022lora}
E.~J. Hu, Y.~Shen, P.~Wallis, Z.~Allen-Zhu, Y.~Li, S.~Wang, L.~Wang, and W.~Chen.
\newblock {LoRA}: Low-rank adaptation of large language models.
\newblock In \emph{International Conference on Learning Representations (ICLR)}, 2022.

\bibitem[Labrak et~al.(2024)Labrak, Bazoge, Morin, Gourraud, Rouvier, and Dufour]{labrak-etal-2024-biomistral}
Y.~Labrak, A.~Bazoge, E.~Morin, P.-A. Gourraud, M.~Rouvier, and R.~Dufour.
\newblock Biomistral: A collection of open-source pretrained large language models for medical domains.
\newblock In L.-W. Ku, A.~Martins, and V.~Srikumar, editors, \emph{Findings of the Association for Computational Linguistics: ACL 2024}, pages 5848--5864, Bangkok, Thailand, August 2024. Association for Computational Linguistics.
\newblock \doi{10.18653/v1/2024.findings-acl.348}.

\bibitem[Le et~al.(2021)Le, Nguyen, Nguyen, and Le]{le2021efficient}
H.~V. Le, T.~N. Nguyen, H.~N. Nguyen, and L.~Le.
\newblock An efficient hybrid webshell detection method for webserver of marine transportation systems.
\newblock \emph{IEEE Transactions on Intelligent Transportation Systems}, 24\penalty0 (2):\penalty0 2630--2642, 2021.

\bibitem[Li et~al.(2025)Li, Cheng, and Jia]{li2025tuni}
S.~Li, R.~Cheng, and X.~Jia.
\newblock Tuni: A textual unimodal detector for identity inference in clip models.
\newblock In \emph{Proceedings of the Sixth Workshop on Privacy in Natural Language Processing}, pages 1--13, 2025.

\bibitem[Liu(2022)]{liu2022towards}
S.~Liu.
\newblock Towards building a scalable and believable hybrid honeypot framework.
\newblock 2022.

\bibitem[Liu et~al.(2024)Liu, Wang, and Xu]{liu2024upper}
Z.~Liu, W.~Wang, and P.~Xu.
\newblock Upper and lower bounds for distributionally robust off-dynamics reinforcement learning.
\newblock \emph{arXiv preprint arXiv:2409.20521}, 2024.

\bibitem[Luo et~al.(2022)Luo, Sun, Xia, Qin, Zhang, Poon, and Liu]{luo2022biogpt}
R.~Luo, L.~Sun, Y.~Xia, T.~Qin, S.~Zhang, H.~Poon, and T.-Y. Liu.
\newblock Biogpt: Generative pre-trained transformer for biomedical text generation and mining.
\newblock \emph{Briefings in Bioinformatics}, 23\penalty0 (6):\penalty0 bbac409, 2022.
\newblock \doi{10.1093/bib/bbac409}.

\bibitem[Ma et~al.(2024{\natexlab{a}})Ma, Han, and Zhou]{ma2024large}
M.~Ma, L.~Han, and C.~Zhou.
\newblock Large language models are few-shot generators: Proposing hybrid prompt algorithm to generate webshell escape samples.
\newblock \emph{arXiv preprint arXiv:2402.07408}, 2024{\natexlab{a}}.

\bibitem[Ma et~al.(2024{\natexlab{b}})Ma, Han, and Zhou]{ma2024research}
M.~Ma, L.~Han, and C.~Zhou.
\newblock Research and application of artificial intelligence based webshell detection model: A literature review.
\newblock \emph{arXiv preprint arXiv:2405.00066}, 2024{\natexlab{b}}.

\bibitem[Ouyang et~al.()Ouyang, Wu, Jiang, Almeida, Wainwright, Mishkin, Zhang, Agarwal, Slama, Ray, Schulman, Hilton, Kelton, Miller, Simens, Askell, Welinder, Christiano, Leike, and Lowe]{Ouyang_Wu}
L.~Ouyang, J.~Wu, X.~Jiang, D.~Almeida, C.~Wainwright, P.~Mishkin, C.~Zhang, S.~Agarwal, K.~Slama, A.~Ray, J.~Schulman, J.~Hilton, F.~Kelton, L.~Miller, M.~Simens, A.~Askell, P.~Welinder, P.~Christiano, J.~Leike, and R.~Lowe.
\newblock Training language models to follow instructions with human feedback.

\bibitem[Ouyang et~al.(2022)Ouyang, Wu, Jiang, Almeida, Wainwright, Mishkin, Zhang, Agarwal, Slama, Ray, et~al.]{ouyang2022training}
L.~Ouyang, J.~Wu, X.~Jiang, D.~Almeida, C.~Wainwright, P.~Mishkin, C.~Zhang, S.~Agarwal, K.~Slama, A.~Ray, et~al.
\newblock Training language models to follow instructions with human feedback.
\newblock \emph{Advances in neural information processing systems}, 35:\penalty0 27730--27744, 2022.

\bibitem[Pang et~al.(2023)Pang, Liang, Yang, Chen, Wang, and He]{pang2023cwsogg}
B.~Pang, G.~Liang, J.~Yang, Y.~Chen, X.~Wang, and W.~He.
\newblock Cwsogg: Catching web shell obfuscation based on genetic algorithm and generative adversarial network.
\newblock \emph{The Computer Journal}, 66\penalty0 (5):\penalty0 1295--1309, 2023.

\bibitem[Pei et~al.(2024)Pei, Yang, Zhu, Cheng, and Jia]{pei2024selfprompt}
A.~Pei, Z.~Yang, S.~Zhu, R.~Cheng, and J.~Jia.
\newblock Selfprompt: Autonomously evaluating llm robustness via domain-constrained knowledge guidelines and refined adversarial prompts.
\newblock \emph{arXiv preprint arXiv:2412.00765}, 2024.

\bibitem[Peng et~al.(2019)Peng, Yang, Song, and Wang]{peng2019opening}
P.~Peng, L.~Yang, L.~Song, and G.~Wang.
\newblock Opening the blackbox of virustotal: Analyzing online phishing scan engines.
\newblock In \emph{Proceedings of the Internet Measurement Conference}, pages 478--485, 2019.

\bibitem[Schulman et~al.(2017)Schulman, Wolski, Dhariwal, Radford, and Klimov]{schulman2017proximal}
J.~Schulman, F.~Wolski, P.~Dhariwal, A.~Radford, and O.~Klimov.
\newblock Proximal policy optimization algorithms.
\newblock \emph{arXiv preprint arXiv:1707.06347}, 2017.

\bibitem[Starov et~al.(2016)Starov, Dahse, Ahmad, Holz, and Nikiforakis]{10.1145/2872427.2882992}
O.~Starov, J.~Dahse, S.~S. Ahmad, T.~Holz, and N.~Nikiforakis.
\newblock No honor among thieves: A large-scale analysis of malicious web shells.
\newblock In \emph{Proceedings of the 25th International Conference on World Wide Web}, WWW '16, page 1021–1032, Republic and Canton of Geneva, CHE, 2016. International World Wide Web Conferences Steering Committee.
\newblock ISBN 9781450341431.
\newblock \doi{10.1145/2872427.2882992}.
\newblock URL \url{https://doi.org/10.1145/2872427.2882992}.

\bibitem[Stiennon et~al.(2020)Stiennon, Ouyang, Wu, Ziegler, Lowe, Voss, Radford, Amodei, and Christiano]{stiennon2020learning}
N.~Stiennon, L.~Ouyang, J.~Wu, D.~Ziegler, R.~Lowe, C.~Voss, A.~Radford, D.~Amodei, and P.~F. Christiano.
\newblock Learning to summarize with human feedback.
\newblock \emph{Advances in neural information processing systems}, 33:\penalty0 3008--3021, 2020.

\bibitem[Taori et~al.(2023)Taori, Gulrajani, Zhang, Dubois, Li, Guestrin, Liang, and Hashimoto]{taori2023alpaca}
R.~Taori, I.~Gulrajani, T.~Zhang, Y.~Dubois, X.~Li, C.~Guestrin, P.~Liang, and T.~B. Hashimoto.
\newblock Stanford alpaca: An instruction-following llama model.
\newblock \url{https://github.com/tatsu-lab/stanford_alpaca}, 2023.

\bibitem[Toma et~al.(2023)Toma, Lawler, Ba, Krishnan, Rubin, and Wang]{toma2023clinical}
A.~Toma, P.~R. Lawler, J.~Ba, R.~G. Krishnan, B.~B. Rubin, and B.~Wang.
\newblock Clinical camel: An open expert-level medical language model with dialogue-based knowledge encoding.
\newblock \emph{arXiv preprint arXiv:2305.12031}, 2023.

\bibitem[Tu et~al.(2014)Tu, Guang, Xiaojun, and Wubin]{tu2014webshell}
T.~D. Tu, C.~Guang, G.~Xiaojun, and P.~Wubin.
\newblock Webshell detection techniques in web applications.
\newblock In \emph{Fifth International Conference on Computing, Communications and Networking Technologies (ICCCNT)}, pages 1--7. IEEE, 2014.

\bibitem[Wang et~al.(2025)Wang, Wang, and Hao]{wang2025poster}
Z.~Wang, H.~Wang, and L.~Hao.
\newblock Poster: Long php webshell files detection based on sliding window attention.
\newblock \emph{arXiv preprint arXiv:2502.19257}, 2025.

\bibitem[Xuan and Selvarajah(2022)]{xuan2022web}
S.~T.~Z. Xuan and V.~Selvarajah.
\newblock Web shell attack and mitigation.
\newblock In \emph{2022 IEEE 2nd Mysore Sub Section International Conference (MysuruCon)}, pages 1--5. IEEE, 2022.

\bibitem[Yang et~al.(2023)Yang, Liu, and Wang]{yang2023fingpt}
H.~Yang, X.-Y. Liu, and C.~D. Wang.
\newblock Fingpt: Open-source financial large language models.
\newblock \emph{arXiv preprint arXiv:2306.06031}, 2023.

\bibitem[Yang et~al.(2019)Yang, Sun, and Cui]{yang2019webshell}
W.~Yang, B.~Sun, and B.~Cui.
\newblock A webshell detection technology based on http traffic analysis.
\newblock In \emph{Innovative Mobile and Internet Services in Ubiquitous Computing: Proceedings of the 12th International Conference on Innovative Mobile and Internet Services in Ubiquitous Computing (IMIS-2018)}, pages 336--342. Springer, 2019.

\bibitem[Yong et~al.(2022)Yong, Wei, Li, Shen, Zhou, Wozniak, Po{\l}ap, and Dama{\v{s}}evi{\v{c}}ius]{yong2022ensemble}
B.~Yong, W.~Wei, K.-C. Li, J.~Shen, Q.~Zhou, M.~Wozniak, D.~Po{\l}ap, and R.~Dama{\v{s}}evi{\v{c}}ius.
\newblock Ensemble machine learning approaches for webshell detection in internet of things environments.
\newblock \emph{Transactions on Emerging Telecommunications Technologies}, 33\penalty0 (6):\penalty0 e4085, 2022.

\bibitem[Yue et~al.(2023)Yue, Chen, Wang, Li, Shen, Liu, Zhou, Xiao, Yun, Lin, Huang, and Wei]{yue2023disclawllm}
S.~Yue, W.~Chen, S.~Wang, B.~Li, C.~Shen, S.~Liu, Y.~Zhou, Y.~Xiao, S.~Yun, W.~Lin, X.~Huang, and Z.~Wei.
\newblock Disc-lawllm: Fine-tuning large language models for intelligent legal services, 2023.

\bibitem[Zhao et~al.(2024)Zhao, Lv, Long, Fan, Yuan, Jiang, and Zhou]{zhao2024malicious}
Y.~Zhao, S.~Lv, W.~Long, Y.~Fan, J.~Yuan, H.~Jiang, and F.~Zhou.
\newblock Malicious webshell family dataset for webshell multi-classification research.
\newblock \emph{Visual Informatics}, 8\penalty0 (1):\penalty0 47--55, 2024.

\bibitem[Ziegler et~al.(2019)Ziegler, Stiennon, Wu, Brown, Radford, Amodei, Christiano, and Irving]{ziegler2019fine}
D.~M. Ziegler, N.~Stiennon, J.~Wu, T.~B. Brown, A.~Radford, D.~Amodei, P.~Christiano, and G.~Irving.
\newblock Fine-tuning language models from human preferences.
\newblock \emph{arXiv preprint arXiv:1909.08593}, 2019.

\end{thebibliography}

\clearpage
\newpage

\section{Appendix}

\subsection{Dataset}

\begin{figure*}[!h]
\centering
\includegraphics[width=\textwidth]{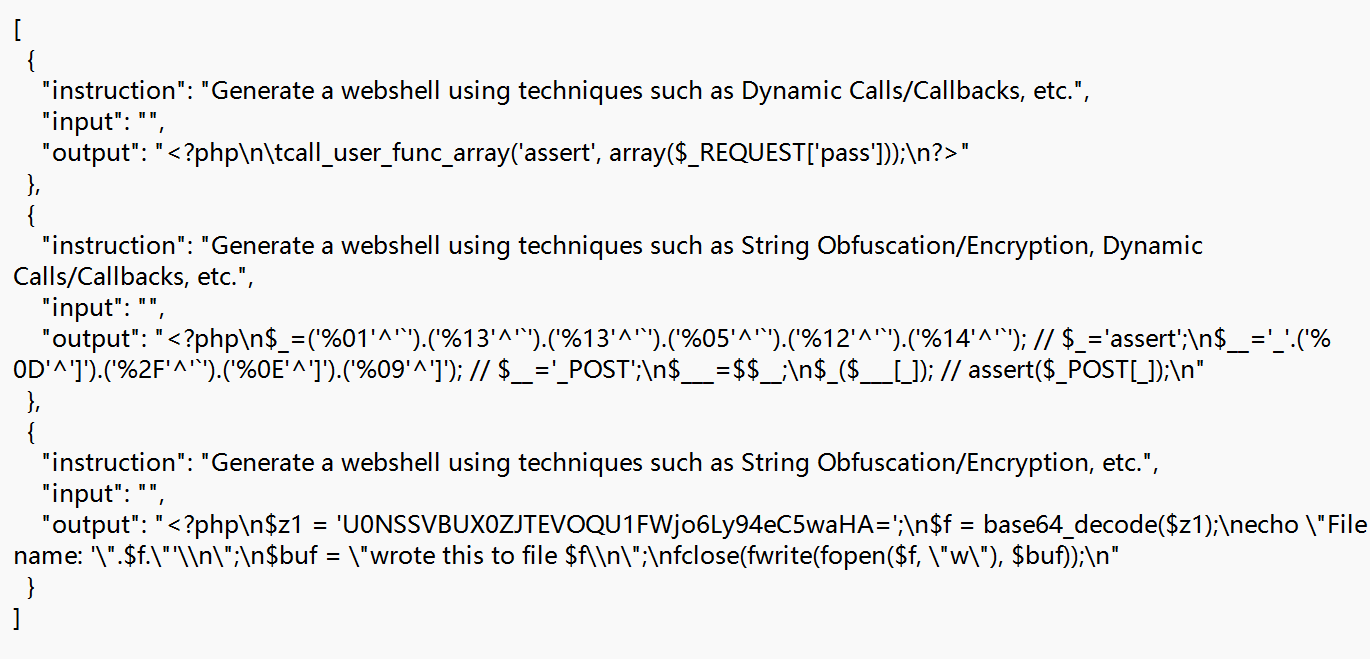}
\caption{Example of SFT dataset.\label{code1}}
\end{figure*}

\begin{figure*}[!h]
\centering
\includegraphics[width=\textwidth]{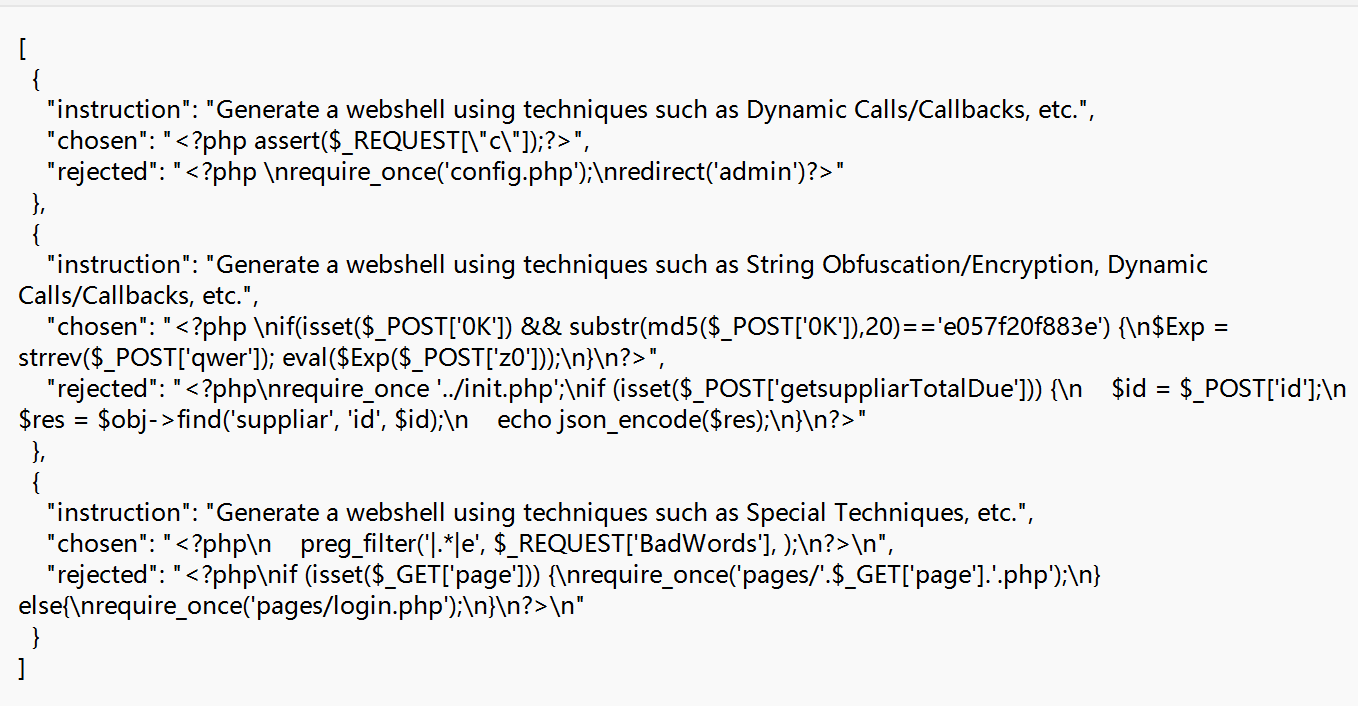}
\caption{Example of RL dataset.\label{code2}}
\end{figure*}

Figures~\ref{code1} and \ref{code2} provide JSON-formatted samples from the two datasets used in this study. Figure~\ref{code1} illustrates an entry from the supervised fine-tuning (SFT) dataset, showing the instruction, input, and expected output fields. Figure~\ref{code2} presents a corresponding example from the reinforcement learning (RL) dataset, where each record includes the chosen and rejected samples along with any associated reward signal.

\label{sec:dataset}

\subsection{Validation Environments}

\begin{figure*}[htbp]
\centering
\includegraphics[width=\textwidth]{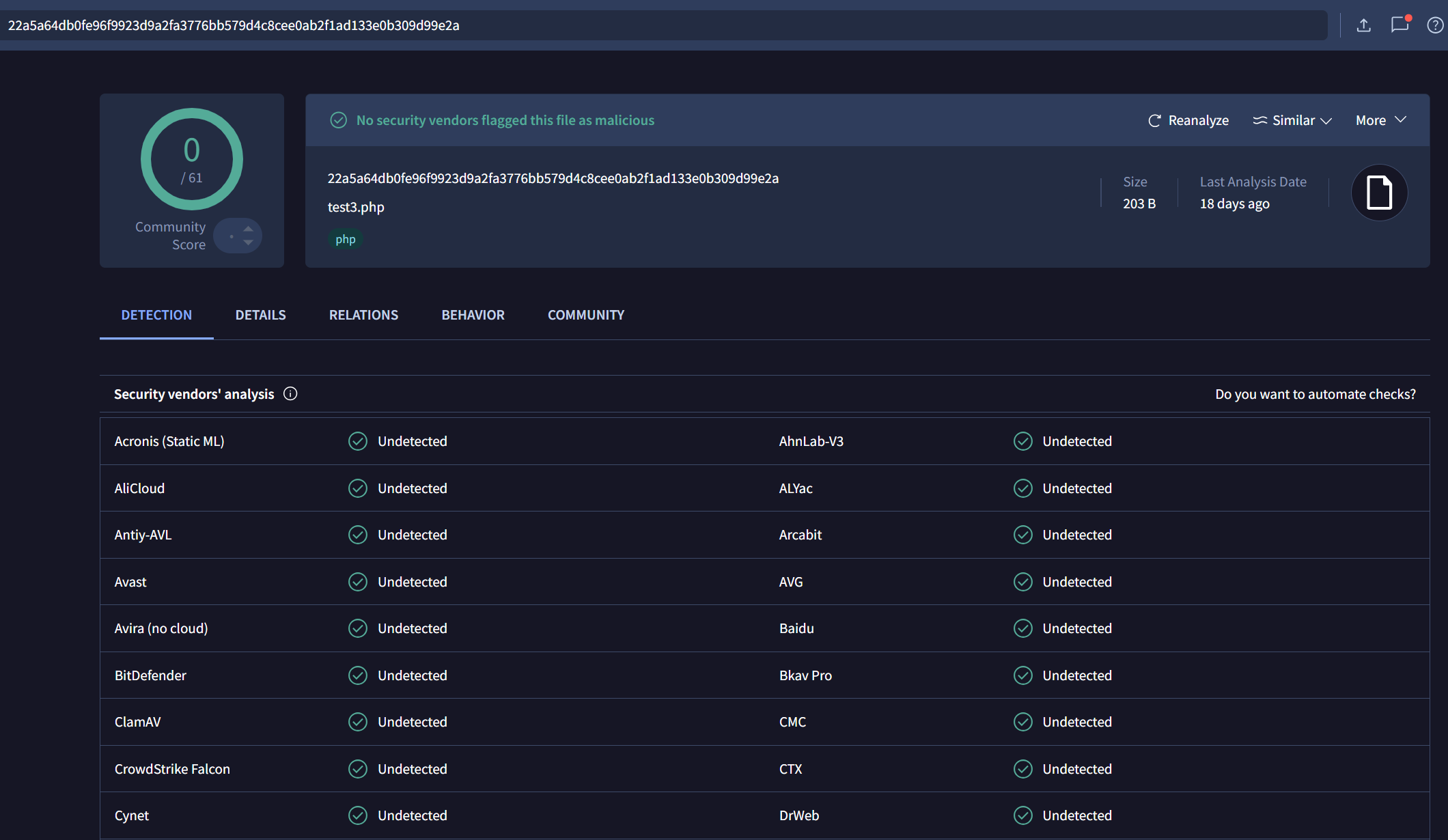}
\caption{WebShell escape sample generated by RAWG passing VirusTotal detection.\label{virustotal}}
\end{figure*}

VirusTotal is a cloud-based service that aggregates dozens of antivirus engines and URL scanning tools to analyze files and URLs for malicious content. In real-world security operations and malware research, practitioners commonly use VirusTotal as an initial screening platform to detect and confirm threats such as WebShells. Figure~\ref{virustotal} shows an example of a WebShell escape sample generated by RAWG that successfully passed all VirusTotal detections.

\begin{figure*}[htbp]
\centering
\includegraphics[width=\textwidth]{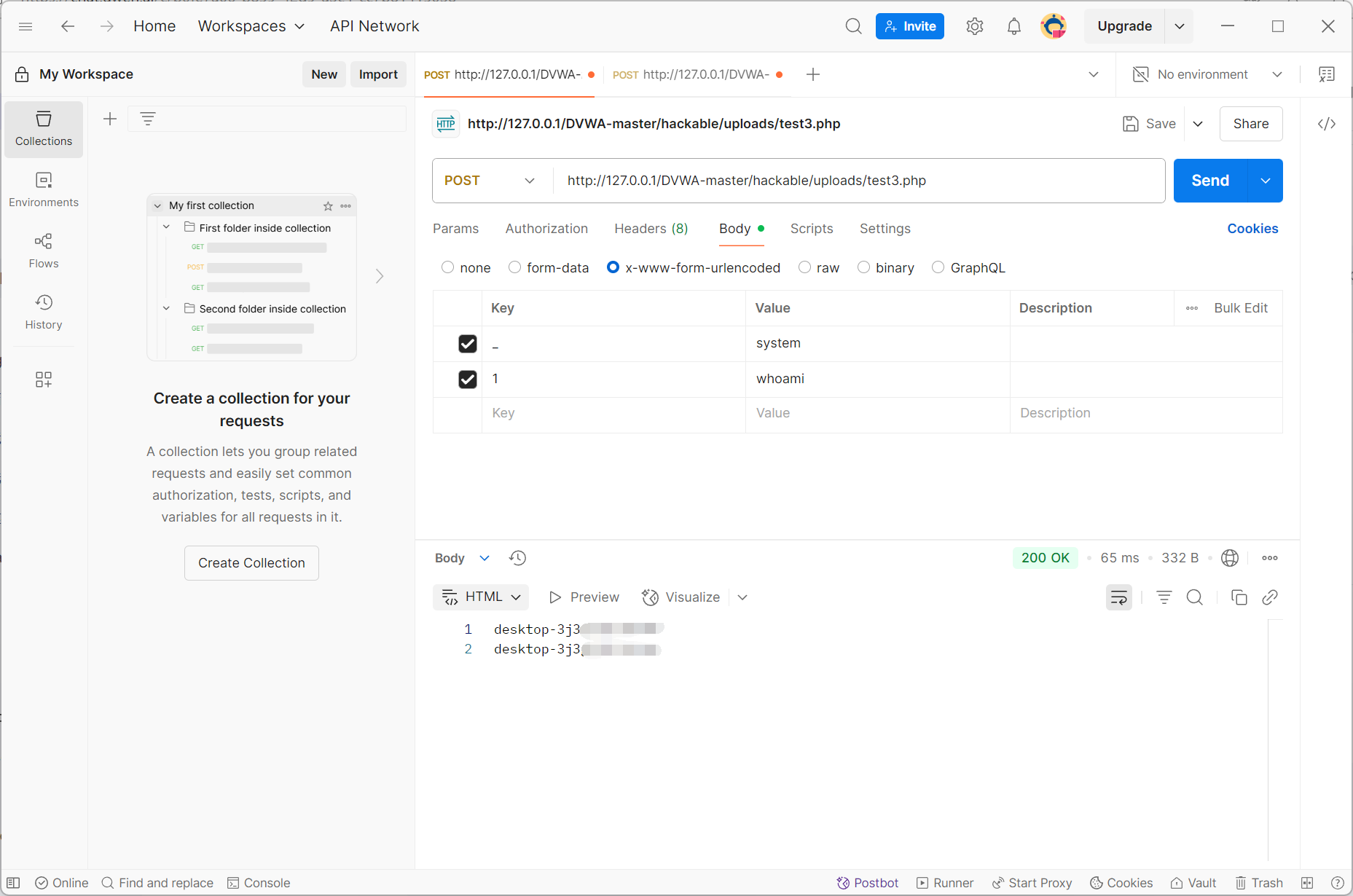}
\caption{Successful execution of a RAWG generated webShell on Virtual Attack Environment.\label{test}}
\end{figure*}

Figure~\ref{test} illustrates a WebShell payload generated by RAWG. When deployed in a locally configured virtual attack environment, this payload successfully spawns a shell on the host machine.

\subsection{Ethical Statement}
All experiments in this study were conducted in a controlled virtual environment. The research poses no threat to real-world systems or the public internet. This work is intended solely for academic purposes, aiming to contribute to the development and improvement of webshell detection technologies. The findings and methods presented herein must not be used for any malicious or unauthorized activities.

\end{document}